\documentstyle[aps,prc,preprint,tighten]{revtex}

\begin{document}
\draft
\title{New technique for phase shift analysis:\\
multi-energy solution of inverse scattering problem}
\author{S.G. Cooper$^{\dag}$, V.I. Kukulin$^{\ddag}$, R.S. Mackintosh$^{\dag}$
 and E.V. Kuznetsova$^{\dag\dag}$}
\address{$^{\dag}$Physics Department, The Open University, Milton Keynes,
 MK7 6AA, U.K.\\$^{\ddag}$Institute of Nuclear Physics, Moscow State 
University, Moscow 119899, Russia.\\
$^{\dag\dag}$Institute of Nuclear Research, Russian Academy of  Sciences, 
Moscow 117312, Russia. }
\date{\today}
\maketitle
\begin{abstract}
We demonstrate a new approach to the analysis of extensive
multi-energy data. For the case of d + $^4$He, we produce a phase
shift analysis covering for the energy range 3 to 11 MeV. The key idea
is the use of a new technique for data-to-potential inversion which
yields potentials which reproduce the data simultaneously over a
range of energies. It thus effectively regularizes the extraction of
phase shifts from diverse, incomplete and possibly somewhat
contradictory data sets.  In doing so, it will provide guidance to
experimentalists as to what further measurements should be made. This
study is limited to vector spin observables and spin-orbit
interactions. We discuss alternative ways in which the
theory can be implemented and which provide insight into the ambiguity
problems. We compare the extrapolation of these solutions to other
energies.  Majorana terms are presented for each potential component.

\end{abstract}

\pagebreak

\setlength{\parindent}{0.3 in}

A well known problem confronting any phase shift analysis (PSA), both
for a single energy and for multiple energies, is the absence of
complete sets of experimental data. A complementary problem is the
occurrence of apparent inconsistencies between data from different
experiments.  These problems are particularly acute for projectiles of
spin $>1/2$.  For example, spin one projectiles require, at each
energy, eight or nine independent measurements (cross sections
$\sigma(\theta)$, vector ${\rm i}\langle
T_{11}(\theta)\rangle$ and tensor $\langle T_{2q}\rangle$
analysing powers, etc.)  Since the PSA solutions based on incomplete
data will be far from unique, we must find a way to apply constraints.
Apart from certain smoothness requirements, it is highly nontrivial to
find {\em general\/} restrictions which are convenient to apply within
the framework of existing PSA methods.
  
In this paper we present a new approach to phase shift analysis, PSA.
In essence, the idea is to find a single multi-component potential to
describe the experimental data over the energy range in question. This
is made possible by a recently developed~\cite{ketal2} direct
data-to-potential inversion technique, the generalized iterative
perturbative method, hereafter GIP, which we describe below. GIP is a
generalization of the established IP $S$-matrix-to-potential inversion
method~\cite{early,ketal1,candm,shirley}.  It enables data for many
energies to be fitted with great computational efficiency by a single
energy dependent potential that is as flexible as required. A PSA
based on a potential, unlike conventional PSAs, leads in a natural way
to sets of phase shifts bearing physically reasonable relationships
between the different partial waves.  Moreover, the model itself will
now reveal any inconsistent data and indicate where new data are
required and thereby be a useful source of experimental guidance. The
GIP potential will have a small energy dependence which, unlike the
rapid energy dependence of the phase shifts, can be compared with the
energy dependence predicted by theory.

We demonstrate our approach by applying the method to 
d + $^4$He and will show that an integrated picture of d + $^4$He
scattering can be obtained from a diverse range of data covering a
substantial energy range. The need for such an approach may be seen
from Ref.~\cite{kuznetsova} where a huge amount of experimental data
for this reaction, including cross section and vector and tensor
analysing powers, are analysed at length by elaborate forms of
PSA. The methods used are described by Krasnopolsky {\em et
al\/}~\cite{kras}. Although the authors of~\cite{kuznetsova} derived
much important information from their PSA (e.g. exact resonance
widths, vertex constant etc.) many results are still on a preliminary
and qualitative level (e.g. the complex tensor mixing parameters,
odd-parity phase shifts). Therefore we believe that the very
considerable experimental effort devoted to this system,
see~\cite{jenny} and many other papers cited in
reference~\cite{kuznetsova}, motivates a new approach. Since d +
$^4$He is a perfect theoretical test case, the rewards will be
physical insight of general relevance to nuclear physics.

In principle the potential searched for should include {\em all\/}
necessary components of nuclear interactions, including central
(Wigner (W) and Majorana (M)) terms, spin-orbit terms (again, both W
and M) and the various possible tensor terms, once more both W and M;
all terms may be complex as required. In determining a suitable
potential, one can impose constraints such as: conformity to known
behaviour of higher partial waves (as in, e.g.~\cite{OPEPtail});
smooth energy dependence of underlying potentials; consistency with
established theories; reproduction of bound- and resonant-state
energies.

In principle it is possible to find a potential fitting data at many
energies by applying standard searching procedures to the parameters
of a sufficiently flexible potential model, whether of standard
multi-parameter or model independent form (e.g. the so-called
Fourier-Bessel analysis). To do this generally entails computationally
expensive and highly non-linear multi-parameter fitting, often leading
to many local minima\cite{greenetal}.  The GIP procedure for direct
data-to-potential inversion solves many of these problems. The
advantages of IP over other methods for $S$-matrix-to-potential
inversion apply here too and are particularly relevant.  The first
advantage is the power to control the exactness of the inversion so
that noisy, incomplete or even partly erroneous data can be fitted
with (one-channel or multi-channel) potentials which do not have
spurious oscillatory features. The second advantage is its virtually
unlimited generalizability. Here we illustrate this feature by
including in our analysis the four Majorana components, normally
omitted in optical model fits.  A further feature of
PSA using the GIP method is its speed and simplicity of application
enabling a thorough exploration of ambiguities. These ambiguities are
{\em not\/} a matter of shallow valley floors in parameter
hyper-space, but appear in the form of apparently disconnected minima.

In the present case we apply the procedure to $S=1$ projectiles,
although for clarity we suppress spin-related subscripts.  The method
involves the following three key elements:\\[3 mm] {\bf (i})\,\,
Expansion of components of the potential (central (c), spin-orbit
(s-o), tensor (t), etc.)  in a suitable basis. For potential component
$k = $c, s-o, t\ldots
\begin{equation}
V^{(k)} = V_0^{(k)} + \sum_j C^{(k)}_j \phi^{(k)}_j(r) \label{first}
\end{equation}
where $C^{(k)}_j$ are coefficients to be determined, $\phi^{(k)}_j(r)$
are the basis functions and $V_0^{(k)}$ is the starting
potential. Note that this expansion applies to both real and imaginary
components and that the notation $\phi^{(k)}_j(r)$ embodies the
possibility that it might be appropriate for different components of
the potential to be expanded in different bases. In particular, real
and imaginary terms, or central and spin-orbit terms, or the Majorana
terms might well require different bases.\\
[2 mm] {\bf (ii)}\,\, The
linear response of the complex $S$-matrix $S_l$ to small changes
$\Delta V(r)$ in the potential:\begin{equation}
\Delta S_l = -\frac{i m }{\hbar^2 k}
\int_0^{\infty} (\psi_l(r))^2 \Delta V(r) {\rm d}r
\label{second}
\end{equation}
with $S_l$ defined in terms of the asymptotic
form of the regular radial wave function as $\psi_l(r) 
\rightarrow I_l(r) - S_l O_l(r)$
where $I_l$ and $O_l$ are incoming and outgoing Coulomb wave functions
of Ref.~\cite{satchlerbook}.  The formulation~\cite{ketal1,ketal2} in
terms of $\delta_l$, where $K_l = \tan{\delta_l}$, is
exactly equivalent. Note that the energy $E_k$ is implicit in
these equations and, for simplicity, we have labelled the channels
only by the orbital angular momentum $l$ although we do include spin
in our calculations.  Equation~\ref{second} can be recast as 
\cite{early,candm}:
\begin{equation}
\frac{\partial S_l}{\partial C_j} =
 -\frac{i m }{\hbar^2 k}\int_0^{\infty} 
(\psi_l(r))^2 \phi_j(r) 
{\rm d}r \label{third} \end{equation} 
where any required superscript $(k)$, 
labelling the potential component, is implicit.\\[2 mm]
{\bf (iii)}\,\, The $\chi^2$ function is defined from:
\begin{equation}
\chi^2 = \sum^N_{k=1} \left(\frac{\sigma_k-\sigma_k^{\rm in}}
{\Delta \sigma_k^{\rm in}} \right)^2 +
\sum_n \sum^M_{k=1} \left(\frac{P_{kn}-P_{kn}^{\rm in}}
{\Delta P_{kn}^{\rm in}} \right)^2 \label{fourth}
\end{equation}
where $\sigma_k^{\rm in}$ and $P_{kn}^{\rm in}$ are the input
experimental values of cross sections and analyzing powers of type $n$
respectively. Since we are fitting data for many energies at once, the
index $k$ indicates the energy as well as angle. 
The data normalising factors can  be introduced as an additional
contribution to Equation~\ref{fourth}.

We must now expand $\chi^2$ in terms of the $C^{(k)}_j$. To do this
we first linearize the
theoretical cross sections and analyzing powers, by
expanding $\sigma_k$ (and $P_{kn}$)  about some current
point $\{C^{(k)}_j(p) \}$:
\begin{equation}
\sigma_k = \sigma_k(C^{(k)}_j(p)) + \sum_{j,l} \left(
\frac{\partial \sigma_k}{\partial S_l(E_k)}\frac{\partial S_l(E_k)}
{\partial C_j^{(k)}}\right)_{C^{(k)}_j(p)}  \Delta C_j^{(k)},
\label{new}
\end{equation}
 which applies at each iterative step $p =0, 1, 2,$\ldots and the
correction (to be determined) for the $j$-th amplitude is $\Delta C_j^{(k)}
=C^{(k)}_j - C^{(k)}_j(p)$.
Equivalent relations are applied for the $P_{kn}$,

Linear equations result from demanding that $\chi^2$ be locally
stationary with respect to variations in the potential coefficients
$C_j^{(k)}$, i.e. the derivatives of $\chi^2$ with respect to the potential
components $C_j^{(k)}$  must vanish.  Solving
these linear equations is straightforward for any reasonable number of
them and yields corrected values $C^{(k)}_j(p)$ [8,10]. We then iterate
the whole procedure, with wave-functions $\psi_l$ in
Equation~\ref{third} calculated using the corrected potentials
from Equation~\ref{first}, until convergence is reached.  This algorithm almost
always converges very rapidly [8,10], in general diverging only when
highly inconsistent or erroneous data have been used or when the
iterative process involves a very unsuitable starting point.\\[1 mm]

Multi-energy inversion is thus reduced to the solution
of simultaneous equations at a series of iterative steps.  To
show how effective this is, we  present the
results of a multi-energy PSA for the d + $^4$He system. For this
initial study, we have selected a small subset of the experimental
data tabulated in \cite{kuznetsova}, in particular the data of Jenny
{\em et al}~\cite{jenny} and that of ~\cite{gruebler,senhouse}.  At
this stage, we have fitted only the cross sections and vector
analysing powers and correspondingly limited ourselves to the
following potential components: Wigner central; Majorana central;
Wigner spin-orbit; Majorana spin-orbit. All terms are complex so that
there are eight components to be determined. The neglect of the
various complex tensor components is justified because their primary
effect is on the tensor analysing
powers. It is well known~\cite{clement,wang} that tensor interactions
in the d + $^4$He system play a moderate role, mainly influencing the
$^3$S$_1$ -- $^3$D$_1$ and $^3$P$_2$ -- $^3$F$_2$ mixing parameters
which are not significant here. The generalisation of GIP to
 yield tensor interactions is under development and we expect a full PSA, 
including all off-diagonal terms, to be presented in due course.
Data renormalization was not considered here since its effect
is small compared to neglect of the tensor force, particularly
for the data sets fitted here~\cite{ketal2,kras}. 

In order to get some understanding of the ambiguity problems, we
consider here two extreme approaches to the fitting process which we
label A and B.  The question of the meaningfulness of the potentials
that are found we leave to later publications.

{\em Approach A\/}\,\, begins the iterative procedure with a starting
potential reflecting very little {\em a priori\/} information
concerning the potential and consists of two components only: simple
real and imaginary central Wigner terms of Gaussian form. The data is
fitted in stages, adding a further potential component at each step
with basis dimensions restricted to two or three Gaussian
functions. Generally convergence results from two or three inversion
iterations at each stage.  By applying a criterion of visual
smoothness, an optimum solution was found, `potential A',
corresponding to $\chi^2/F =18.7$.  Fits giving a lower $\chi^2/F$ are
possible with a larger basis, but the corresponding $|S|$ also show a
significant unitarity breaking for certain $l,j$. This case
involves about 20 independent parameters.

{\em Approach B\/}\,\,  starts the iterative procedure with
a potential derived by inversion~\cite{npa625} of $S_{lj}$ from the 
multi-configuration RGM calculations of Kanada {\em et al\/}~\cite{kkst}  
which include S-wave deuteron breakup. This approach gave `potential B'
with $\chi^2/F= 5.84$ but is accompanied by a significant breaking
of unitarity in the S wave. (The results are described in detail
in Ref.~\cite{Puri}.) 

In both approaches energy dependence is included only in the imaginary
components.  The procedure used follows Ref.~\cite{candm2}, which
applies for shape invariant energy dependent potentials.  Since the
inelastic threshold is at $E_{\rm th}= 3.3$ MeV, we expect the
imaginary components to increase rapidly as the energy rises above
$E_{\rm th}$ and so we assume that all parts of the imaginary
potential increase linearly with $(E-E_{\rm th})$. In fact, the
results are insensitive to this energy dependence. Both the
detailed form of the imaginary potentials and the imaginary phase
shifts are less well determined than the corresponding real quantities
and qualitative features of the data can be reproduced with a real
potential alone.

In Figure 1 we display, for representative energies over the complete
energy range of 3 -- 11.5 MeV, typical fits to cross
sections~\cite{senhouse} and in Figure 2, analyzing
powers~\cite{gruebler}.  Both $\sigma(\theta)$ and ${\rm i}\langle
T_{11}(\theta)\rangle$ are very well fitted over the entire energy
range. Closely compatible fits to the data of Ref.~\cite{jenny} were
found, both visually and in the values of $\chi^2$. 
All the quoted $\chi^2/F$ values
apply to the fit over the full energy range, but are only relative 
since the tabulated data did not include all the sources of error
discussed in the original papers. We have found
that although the contribution of the mixing parameters
to the cross-section is almost negligible, there is a more noticeable
effect on the fit to  ${\rm i}\langle
T_{11}(\theta)\rangle$.

The bound state energy of the $^4$He -- d system, which can be
identified as the ground state energy of $^6$Li in the $^4$He -- d
channel, is not included in these inversions.  Potential A gives
$E_{\rm B}= - 2.26$ MeV ($E_{\rm B}^{\rm expt} = - 1.472$ MeV).  Note
that this energy is extremely sensitive to the form of the potentials
and to the energy dependence of the d -- $^4$He $^3$S$_1$ phase
shifts~\cite{blokh}.

In Figure 3 we present the real parts of potentials A.
Known ambiguity problems suggest this potential is
almost certainly not unique.  Within either approach, A or B, certain
potential components are more reliably determined than others, the
real central Wigner term being the best determined. Its volume
integral is consistent with global potentials and also with volume
integrals of the corresponding potential derived by 
$S$-matrix to potential inversion for the theoretical $S_l$ of resonating
group model (RGM) calculations~\cite{ketal2,npa625,blokh,kuk-new}.

The phase shifts corresponding to the solution A for $l\le 4$ are
displayed in Figure 4 for an energy range of 0 to 15 MeV laboratory
energy, i.e. extrapolating outside the range of the data.  This figure
also includes the results of a previous analysis~\cite{kuk-new}. The
really difficult problem for all previous (standard) PSAs was to
achieve a low energy description of odd partial waves (i.e. $^3$P$_j$
with $j=0, 1, 2$ and $^3$F$_j$ with $j=2, 3, 4$), due to the weak
sensitivity of cross sections and analysing powers to the odd partial
waves ~\cite{kuznetsova}. Thus, by fitting all significant partial
waves independently in the course of a standard
PSA~\cite{kuznetsova,jenny}, a range of solutions are possible which
are consistent with the data. The resulting odd-parity phase shifts
have very large error bars.  In the present method for phase shift
analysis a further restriction is applied by demanding a smooth
underlying potential and therefore the approach should lead, in
principle, to much more reliable and accurate values for all phase
shifts than found in previous PSAs~\cite{kuznetsova,jenny}.

The comparison in Fig.~4 of our new PSA solution with previous
results, shows that the agreement for even partial waves is quite
close while there is less agreement for odd partial waves.  This is
probably due our neglect of tensor forces.  Reliable knowledge of the
odd partial wave phase shifts is crucially
important~\cite{kuk-new,kuk-pom}, since the nature of the deuteron --
nucleus interaction, particularly for d + $^4$He, is different for
even and odd partial waves. The even parity d + $^4$He interaction is
determined by an intermediate state in which two nucleons in the
incident deuteron occupy two (1p)-orbitals beyond the $^4$He
core. However, for odd parity, the two outer nucleons occupy
non-overlapping 1p--2s or 1p--2d orbits (designating orbits $Nl$, with
$N$ the number of oscillator quanta). Thus, since the N--N interaction
is short ranged compared to the range of d -- $^4$He interaction, the
contribution of virtual breakup should be higher for odd than for even
partial waves and the sensitivity to the N + $\alpha$ interaction
should also be higher. Due to this feature of the d + $^4$He
interaction, the p- and f-wave phase shifts have been
shown~\cite{kuk-pom} to give a strong test of supersymmetrical aspects
of composite particle interactions and the structure of tensor
interactions of deuterons.  A further step now is to include in our
potential terms which have never previously been considered for
nucleus-nucleus interactions: complex Majorana tensor
forces. Preliminary results~\cite{wip} show that the Majorana tensor
force is approximately as strong as the Wigner tensor force.

In summary: we have demonstrated a new approach to PSA based
on a linearized iterative approach to
direct inversion from multi-energy data to potentials. The example
presented, approach A,  involved far fewer parameters than 
a conventional PSA (about a hundred for this case). 
The new method is computationally efficient and avoids
many drawbacks and instabilities of conventional PSAs, especially in
cases of projectile of spin 1 or greater when one generally has an
incomplete data set with data at many relevant energies absent or
having large error bars.  As well as correct phase shifts, the
potential itself is of great interest since it can be used as input
for other calculations and can also be compared with potentials found
by double folding procedures or by inversion from $S_l$ obtained from
RGM and other theoretical models.
  
\section*{Acknowledgements} The authors are very grateful to  V.N.
Pomerantsev for many fruitful discussions on the topics discussed
here.  One of the authors, (V.I.K.) is grateful to Willi
Gruebler for supplying him the full tables of experimental data of the
Z\"urich group.  We are also most grateful to the UK EPSRC for grant
GR/H00895 supporting S.G. Cooper, the Russian Foundation for Basic
Research (grant 97-02-17265) for financial assistance and to the Royal
society (UK) for supporting a visit by V.I. Kukulin to
England.

\setlength{\parindent}{0.0 in}

\newpage
\begin{figure}
\caption{For deuterons scattering from $^4$He, fits to differential cross
sections of Senhouse and Tombrello at selected energies. The solid line
is the fit for potential A, the dashed line for potential B.}
\end{figure}

\begin{figure}
\caption{For deuterons scattering from $^4$He, fits to vector analysing power
data  of Gruebler {\em et al\/} at selected energies. The solid line is
the fit for potential A, the dashed line for potential B.}
\end{figure}

\begin{figure}
\caption{The real parts of potential A. 
From top, the Wigner central and spin-orbit, then the Majorana central and 
spin-orbit.}
\end{figure}

\begin{figure}
\caption{The real phase shifts for fit A (solid line) compared with 
the results of a conventional phase shift analysis (filled circles).}
\end{figure}

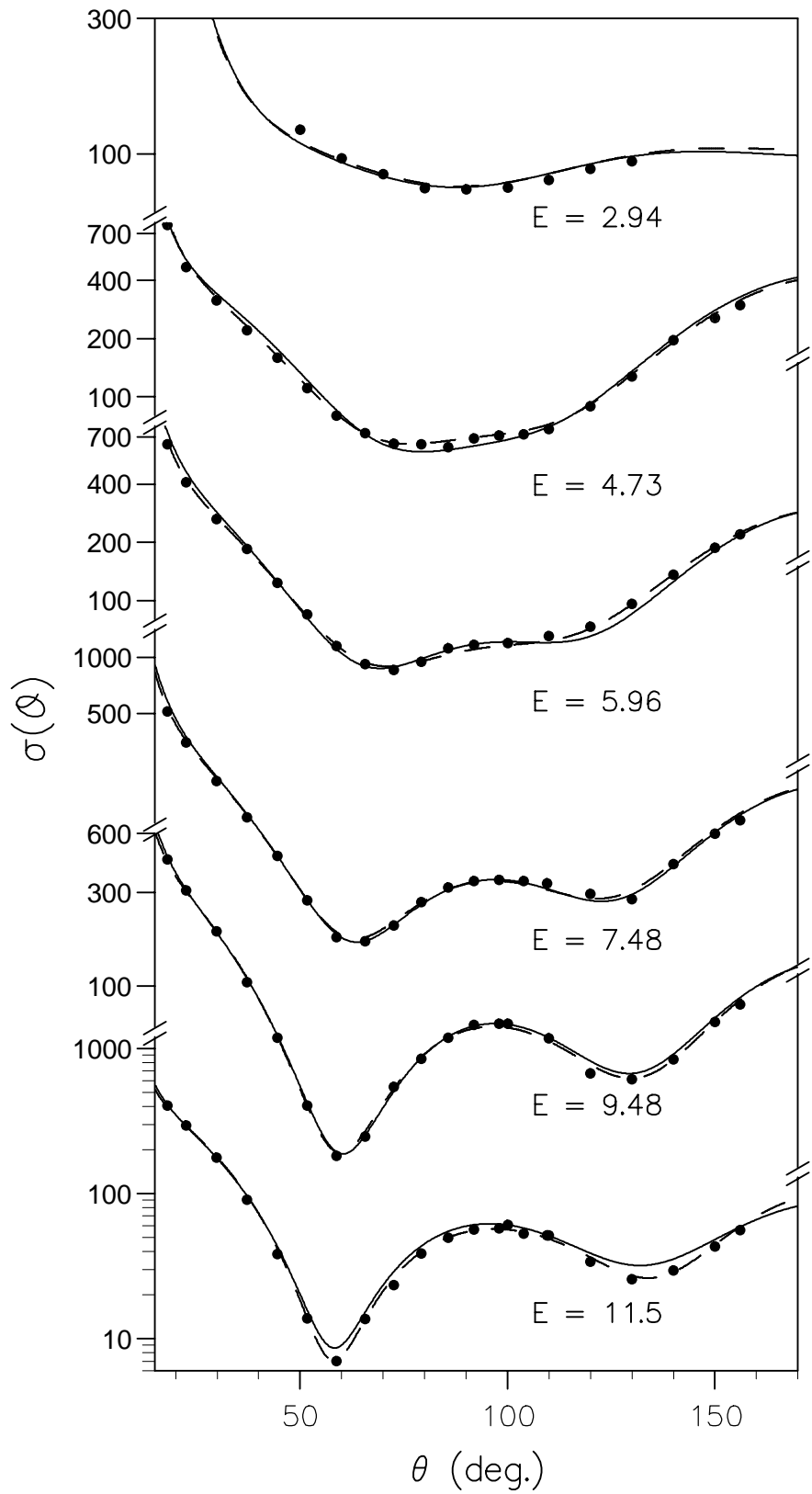
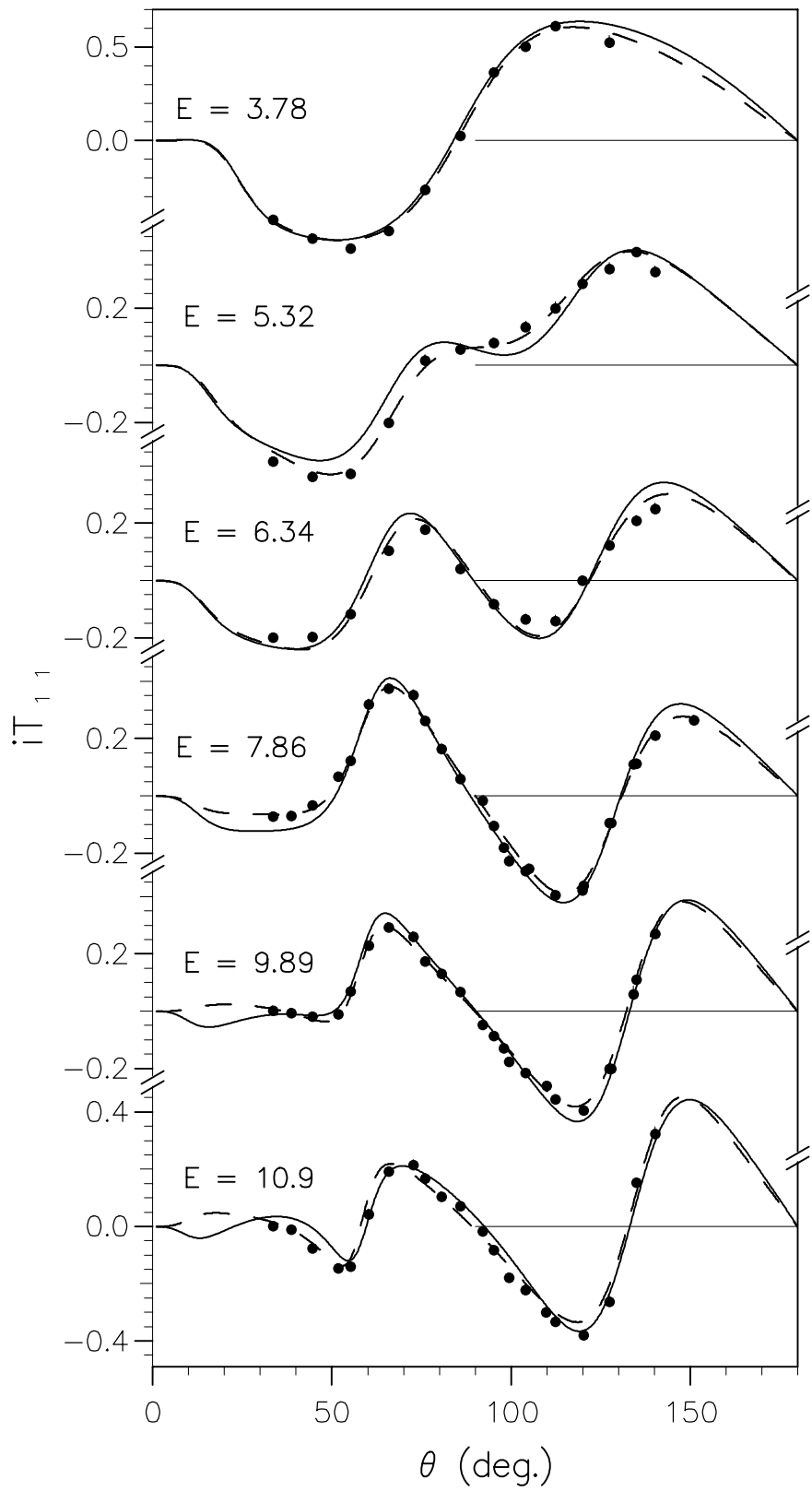
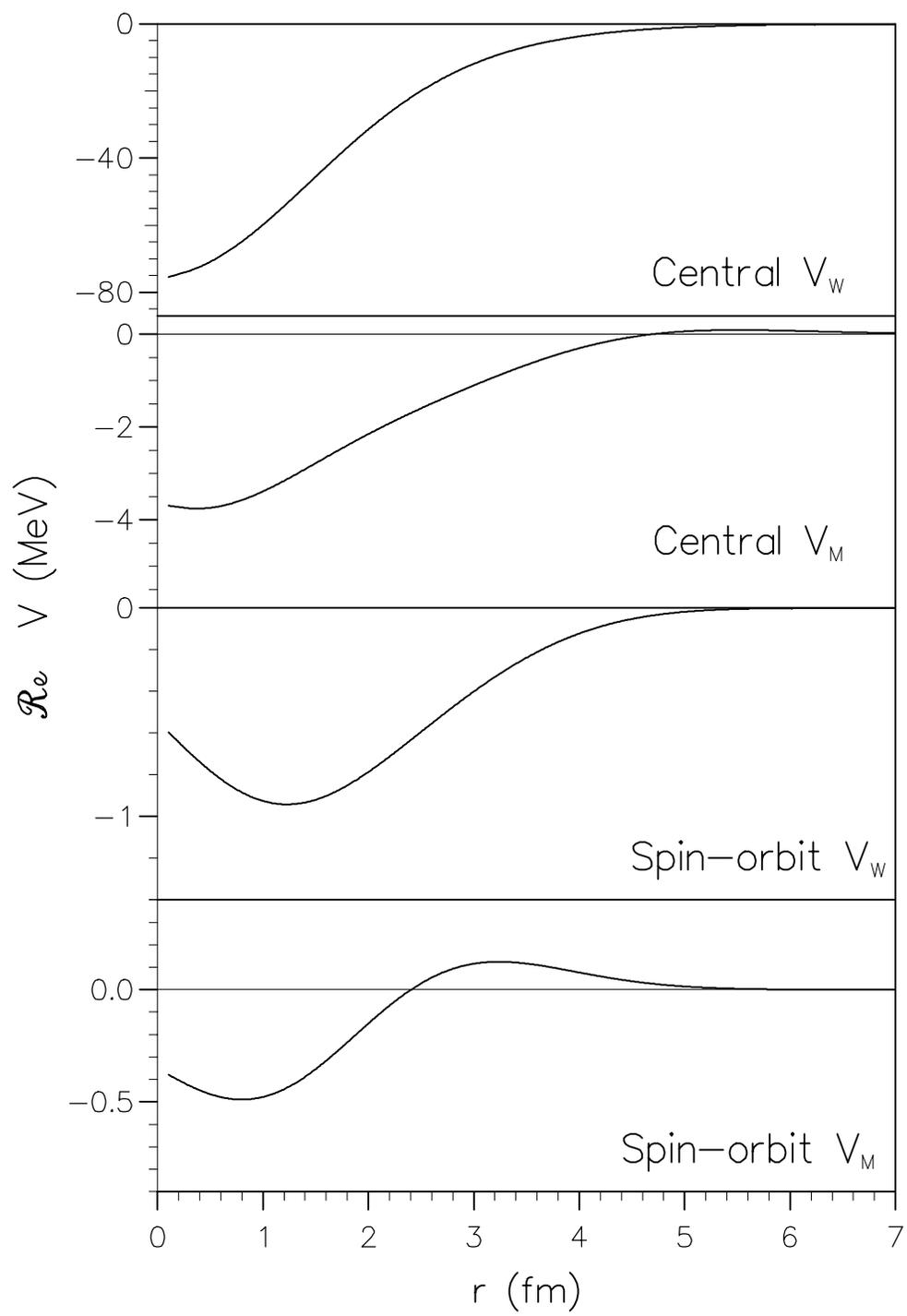
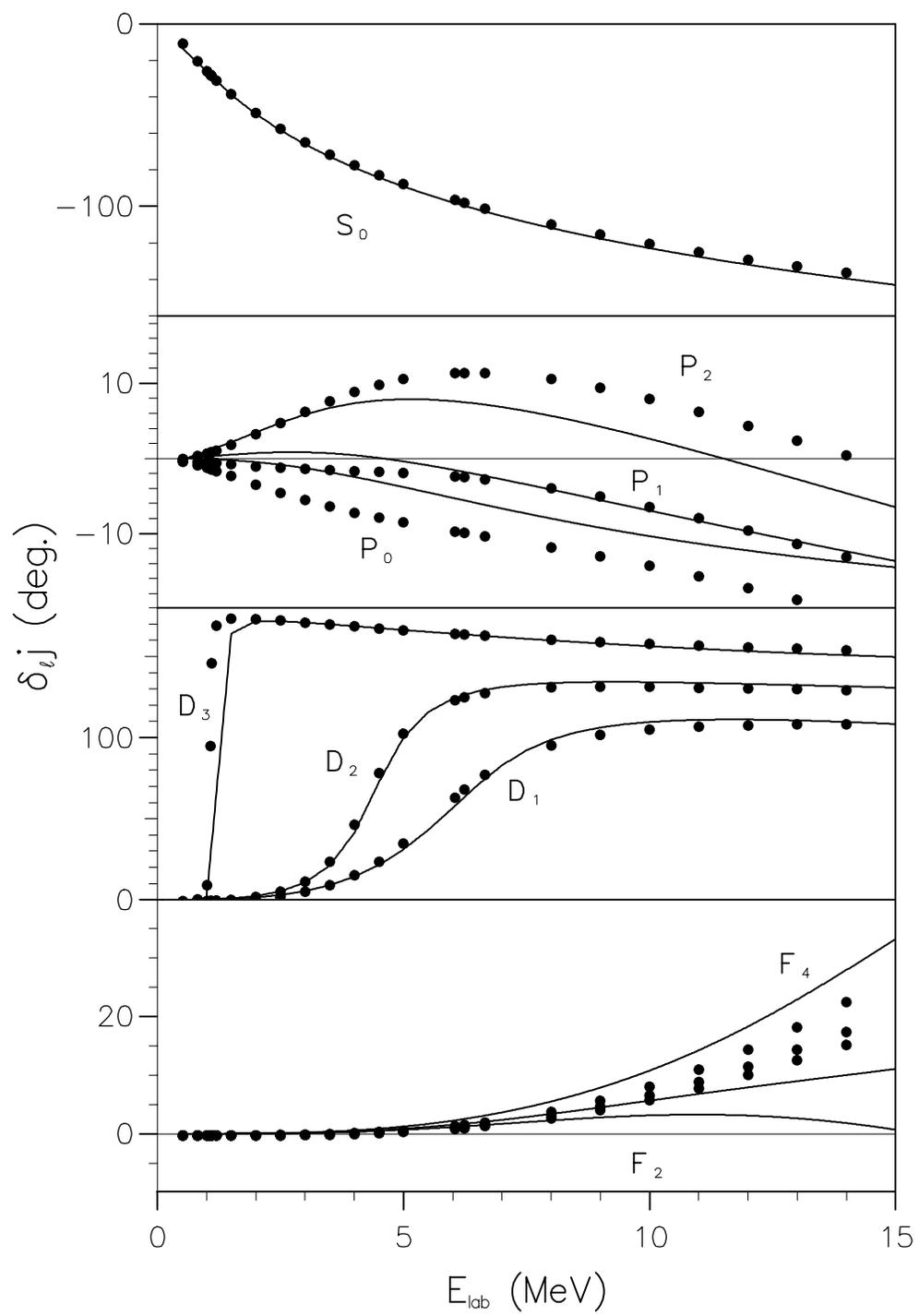
\end{document}